\documentclass[a4paper,11pt]{article}
\usepackage{pos}

\title{Automatic data processing for Baikal-GVD neutrino observatory}

\author[a]{V.A.~Allakhverdyan}
\author[b]{A.D.~Avrorin}
\author[b]{A.V.~Avrorin}
\author[b]{V.M.~Aynutdinov}
\author[c]{R.~Bannasch}
\author[d]{Z.~Barda\v{c}ov\'{a}}
\author[a]{I.A.~Belolaptikov}
\author[a]{I.V.~Borina}
\author[a,1]{V.B.~Brudanin}
\author[e]{N.M.~Budnev}
\author[a]{V.Y.~Dik}
\author[b]{G.V.~Domogatsky}
\author[b]{A.A.~Doroshenko}
\author[a,d]{R.~Dvornick\'{y}}
\author[e]{A.N.~Dyachok}
\author[b]{Zh.-A.M.~Dzhilkibaev}
\author[d]{E.~Eckerov\'{a}}
\author[a]{T.V.~Elzhov}
\author[f]{L.~Fajt}
\author[g,1]{S.V.~Fialkovski}
\author[e]{A.R.~Gafarov}
\author[b]{K.V.~Golubkov}
\author[a]{N.S.~Gorshkov}
\author[e]{T.I.~Gress}
\author[a]{M.S.~Katulin}
\author[c]{K.G.~Kebkal}
\author[c]{O.G.~Kebkal}
\author[a]{E.V.~Khramov}
\author[a]{M.M.~Kolbin}
\author[a]{K.V.~Konischev}
\author[h]{K.A.~Kopa\'{n}ski}
\author[a]{A.V.~Korobchenko}
\author[b]{A.P.~Koshechkin}
\author[i]{V.A.~Kozhin}
\author[a]{M.V.~Kruglov}
\author[b]{M.K.~Kryukov}
\author[g]{V.F.~Kulepov}
\author[h]{Pa.~Malecki}
\author[a]{Y.M.~Malyshkin}
\author[b]{M.B.~Milenin}
\author[e]{R.R.~Mirgazov}
\author[a]{D.V.~Naumov}
\author[a]{V.~Nazari}
\author[h]{W.~Noga}
\author[b]{D.P.~Petukhov}
\author[a]{E.N.~Pliskovsky}
\author[j]{M.I.~Rozanov}
\author[a]{V.D.~Rushay}
\author[e]{E.V.~Ryabov}
\author[b]{G.B.~Safronov}
\author*[a]{B.A.~Shaybonov}
\author[b]{M.D.~Shelepov}
\author[a,d,f]{F.~\v{S}imkovic}
\author[a]{A.E. Sirenko}
\author[i]{A.V.~Skurikhin}
\author[a]{A.G.~Solovjev}
\author[a]{M.N.~Sorokovikov}
\author[f]{I.~\v{S}tekl}
\author[b]{A.P.~Stromakov}
\author[a]{E.O.~Sushenok}
\author[b]{O.V.~Suvorova}
\author[e]{V.A.~Tabolenko}
\author[e]{B.A.~Tarashansky}
\author[a]{Y.V.~Yablokova}
\author[c]{S.A.~Yakovlev}
\author[b]{D.N.~Zaborov}

\affiliation[a]{Joint Institute for Nuclear Research, Dubna, Russia}
\affiliation[b]{Institute for Nuclear Research, Russian Academy of Sciences, Moscow, Russia}
\affiliation[c]{EvoLogics GmbH, Berlin, Germany}
\affiliation[d]{Comenius University, Bratislava, Slovakia}
\affiliation[e]{Irkutsk State University, Irkutsk, Russia}
\affiliation[f]{Czech Technical University in Prague, Prague, Czech Republic}
\affiliation[g]{Nizhny Novgorod State Technical University, Nizhny Novgorod, Russia}
\affiliation[h]{Institute of Nuclear Physics of Polish Academy of Sciences (IFJ~PAN), Krak\'{o}w, Poland}
\affiliation[i]{Skobeltsyn Institute of Nuclear Physics, Moscow State University, Moscow, Russia}
\affiliation[j]{St.~Petersburg State Marine Technical University, St.Petersburg, Russia}

\note{Deceased.}

\emailAdd{bair@jinr.ru}

\abstract{Baikal-GVD is a gigaton-scale neutrino observatory under construction in Lake Baikal. 
It currently produces about 100 GB of data every day. For their automatic processing, 
the Baikal Analysis and Reconstruction software (BARS) was developed. 
At the moment, it includes such stages as hit extraction from PMT waveforms, assembling events from raw data, 
assigning timestamps to events, determining the position of the optical modules using an acoustic positioning system, 
data quality monitoring, muon track and cascade reconstruction, as well as the alert signal generation. 
These stages are implemented as C++ programs which are executed sequentially one after another and can be represented 
as a directed acyclic graph. The most resource-consuming programs run in parallel to speed up processing. 
A separate Python package based on the luigi package is responsible for program execution control. 
Additional information such as the program execution status and run metadata are saved into a central database 
and then displayed on the dashboard. Results can be obtained several hours after the run completion.}

\FullConference{37$^{\rm{th}}$ International Cosmic Ray Conference (ICRC 2021)\\
		July 12th -- 23rd, 2021\\
		Online -- Berlin, Germany}

\begin{document}
\maketitle

\section{Introduction}

The Baikal-GVD underwater neutrino telescope is a cubic kilometer scale detector being constructed in Lake Baikal, 
about 100 km from Irkutsk, Russia \cite{Belolap}. 
It will be a multifunctional device with the main purpose to measure high-energy cosmic neutrinos with high angular 
precision and with high sensitivity for astrophysical neutrino sources located in the Southern Hemisphere. 
Detection of high-energy cosmic neutrino sources will shed light to the long standing puzzle of the origin of cosmic rays.  
The telescope is a three dimensional array of photodetectors located underwater at the depths of 750 - 1270 m and about 4 km offshore. 
The photodetector is 10-inch Hamamatsu PMT housed in a pressure-resistant glass sphere that is called an optical module (OM). 
The OMs are attached to the vertical stainless steel ropes that are called strings and register Cherenkov light from 
the secondary charged particles produced in neutrino interactions. 
Baikal deep water is characterised by high optical transparency and low light scattering \cite{Ryabov}. 

The detector is segmented into clusters of 8 strings - one central and 7 peripheral strings at a radius of about 60 m. 
Each string consists of 36 optical modules. Eight clusters have been installed so far with 2014 OMs in total and are currently taking data. 
Distance between central strings of the clusters is about 300 m except for cluster 8 that is located just 200 m away from the nearest clusters. 
Each cluster has its own lakebed cable communication and power line to the shore, DAQ and represents as a standalone detector.

A cluster produces about 12-16 GB of raw data for typical 24 hours run. 
The data are compressed by factor of 4. Typical trigger rate of the cluster is about 40-50 Hz. 
The total number of raw data files per run is about 100, approximately 40 MB per compressed file. 
But during the period of increased water chemiluminescence, the rates can increase by a factor of a few \cite{Rastislav}. 
Based on the first 5 years of data taking with the partially completed detector several 
neutrino candidate events of potentially astrophysical origin were observed \cite{Zhan}.

\section{Raw data movement and storage}

The OMs in a string are grouped into three sections. There are 3 sections in a string. 
Each section consists of 12 OMs and a Central module (CM) that digitizes signals from them 
in 200 MHz FADC and realizes trigger logic \cite{Aynutdin}. 
Additionaly, there are acoustic positioning system data, OM monitoring data and White-Rabbit synchronisation data.
Raw data is firstly transferred to the shore by bed fibre-optic cable line and stored on the onsite computer farm. 
Then data files as soon as they appeared on the shore are transferred immediately to the central storage and 
processing facility in Joint Institute for Nuclear Research (JINR) in Dubna, Russia through the 300 Mbit/s radio link 
over the lake to Baikalsk town and then through the Internet. 
Each raw file that has several minutes exposure PMT waveforms appears at the central storage with a latency of less than a minute. 
Technically raw data are transferred by rsync file-copying tool. 
To minimize a latency new raw data files are stored in a buffer where rsync transfers them to Dubna 
and removes them after successful operation. 

Raw and processed data are stored in the EOS storage system. 
To use a benefit of native EOS copying tools raw data are collected first to the relay linux machine and then are moved to the EOS. 
When a run is finished an empty so called 'end' file that contains the number of the last raw file in a filename is created by DAQ. 
When it appears in the central storage an appropriate record in the central database that is located in JINR 
is created signaling that the raw data is ready to be processed. Such simple custom-made approach provides the reliable 
data movement due to the absence of external dependencies.

\section{Quasi online software}

Baikal-GVD processing software consists of core and managing parts. 
The former is a set of C++ programs written on the basis of BARS (Baikal Analysis and Reconstruction Software) 
framework that realises specific processing algorithms which take raw or 
derived data as an input and produce other derived data as an output \cite{BARS}. 
The processing algorithms are encapsulated in different programs to achieve flexibility of the system, 
easier maintenance, upgrade and adding more functionality. 
The output data from one program are the input for others. 
So a set of programs form a directed acyclic graph. 
At the moment the graph has all main elements to obtain ready-for-physics analysis data and for producing alerts.
The second part is the PyBARS workflow management system that realises an execution of the graph and will be described later.

\subsection{Processing chain}

\begin{figure}
	\includegraphics[width=0.9\linewidth]{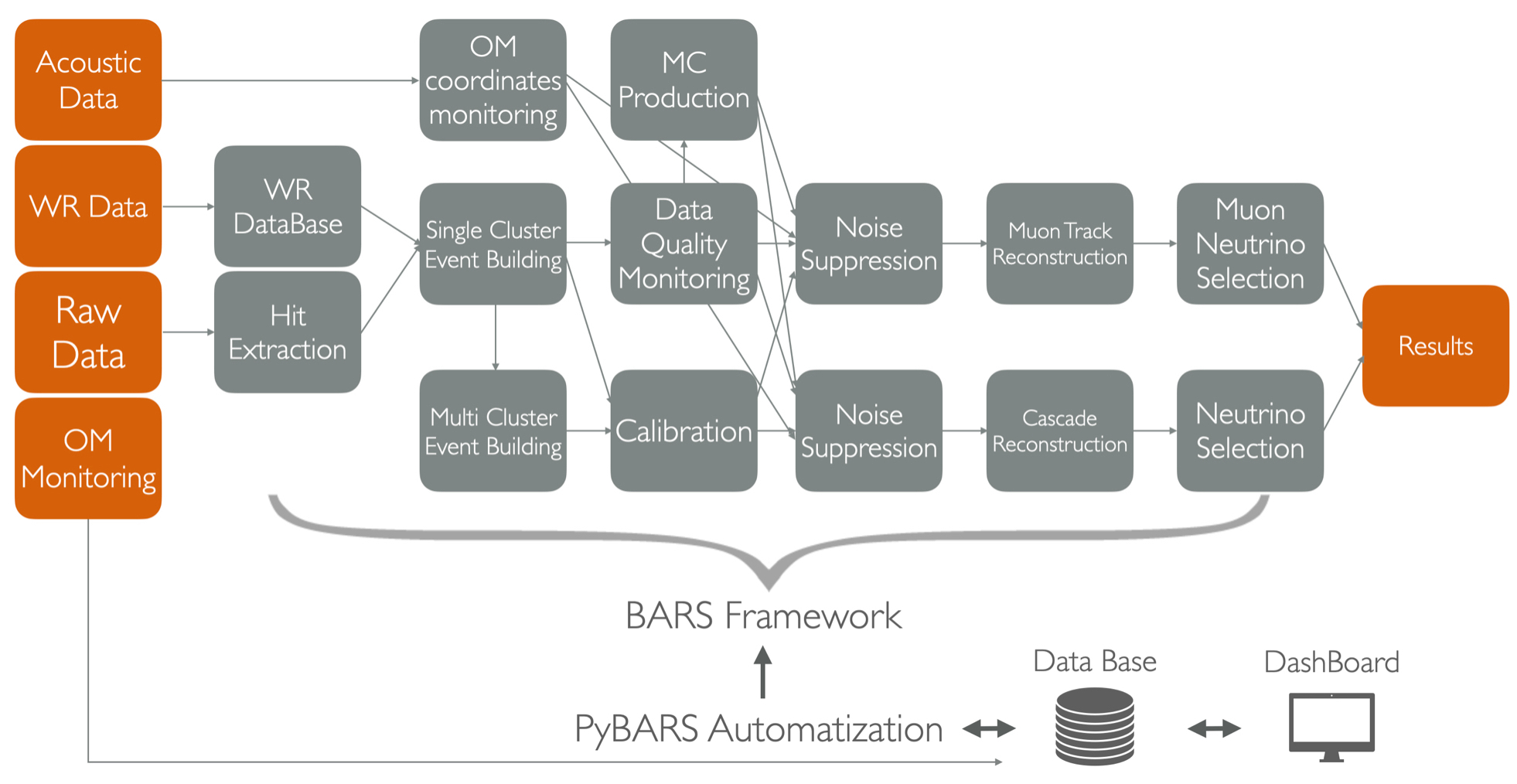}
	\caption{Overview of the data processing in Baikal-GVD. Arrows show data flow.}
	\label{processing_scheme}
\end{figure}

The data processing chain is shown in Fig. \ref{processing_scheme}.
The first step of data processing is to extract main characteristics from PMT pulse waveforms grouped by 12 OMs in a section. 
The waveforms are obtained by CM firmware in such a way that there are 10 pedestal FADC counts before the pulses which are used to subtract it. 
First several tests are applied to waveform to be sure that it is a real signal rather than electronics cross-talk. 
Using piecewise-linear approximation of the waveform the signal arrival time (time mark) is calculated at 
the front at half maximum level and later it is shifted using time walk correction function that was obtained from time calibration. 
Also the maximum amplitude, the time over half maximum and the charge as a sum of digitized counts are calculated. 

There are two synchronization systems embedded in the cluster architecture. 
One is the native system developed by the collaboration using the similar electronics as in section, 
and produce data in the same format as sections, synchronised with world time with 5 ns precision. 
Another third party White - Rabbit system produces data as text event time series with a precision of 1 ns. 
They are accumulated in Influx database that is convenient for such kind of data and for subsequent processing.
Since hits are extracted from the sections and event time stamps from the two synchronisation systems 
are prepared then all these data are combined to single cluster events. 
Next, events on different clusters that occurred simultaneously are merged to the multi-cluster events. 
At the moment this step is not embedded to the quasi online processing chain and is run separately. 

The GVD acoustic data is processed and collected independently.
At regular periods the shore positioning software reconstructs coordinates of acoustic beacons installed 
along the strings and stores it locally.
This data is synchronised to the MariaDB database in JINR cloud every 2 minutes.
At JINR the beacon coordinates for current season undergoes daily automated post-processing. 
The coordinates are fetched from the database, cleaned from outliers, and then linearly 
interpolated at 2 minute intervals, so that each interval 
would have a full set of interpolated beacon coordinates.
The resulting coordinates are then exported to a variety of formats used by different reconstruction pipelines.
For the reconstruction the interpolated beacon positions are used to interpolate OM positions 
along the string based on the string layout and piecewise linear string model.
The OM coordinates are reconstructed on event per event basis.

Next, data quality monitoring (DQM) validates the collected data, verifies the detector status, 
produces amplitude calibration constants and determines the current lake noise level \cite{Sorokovikov}. 
Time calibration is carried out in the separate calibration runs that are obtained once a week with 
switched on artificial light sources \cite{Lukas}. 
The latest calibration constants are applied to the single cluster events at this stage. 
Because of all the necessary information about current run is obtained we can produce 
run-by-run MC simulated events from atmospheric muons and neutrinos fast using prepared in advance intermediate 
MC data that are obtained without noise, trigger condition, dead channel configuration. 
From this moment, the MC data can be processed in parallel with the real data. 
MC and real events are stored in the unified event format in order to apply for them the same processing code further. 
Next, the fast muon track reconstruction algorithm as well as two versions of cascade reconstruction algorithms 
with their specific noise hit suppression techniques are applied. 
If specific selections on quality upgoing reconstructed events are fulfilled, the internal alert is produced and sent to the mailing list.  

\subsection{Workflow management}

PyBARS python code based on the luigi python package \cite{luigi} manages  execution of the processing programs in a pre-defined order. 
It starts to build the graph of programs from the end. For now, it is neutrino selection elements as shown in Fig. 1. 
Each processing program in order to be executed requires the presence of input data files which are the output of other programs. 
Full input and  output file names and their paths for each program are standardised. 
Each program has its own python counterpart (envelope) which defines dependencies from the other programs. 
The dependencies are resolved in the following. Each processing program has a '--promise' argument. 
Program execution with this argument just only prints full output file names. 
If output files of the required programs exist the considered program will be executed. 
In the absence of output files, the required programs have to be executed with the requirements resolved. 
Thus the graph of programs are built up to raw data files. PyBARS code has an internal mechanism to 
do a parallel execution of the programs on the event basis as well as on CM basis. 
It can execute multiple copies of a program with different event or CM ranges and join the multiple 
results by execution of a program with additional --collect argument which triggers the appropriate algorithm inside the program. 
For different types of runs such as calibration runs the graph can be easily rebuilt by requirement of different programs as the end ones.

\section{Data processing and monitoring}

There are 10 virtual machines with 28-32 CPU cores, each are used for the quasi online data processing. 
PyBARS workflow runs permanently on them using the latest approved calibration constants. 
As soon as full set of run's raw data files reaches JINR storage, PyBARS workflow is triggered by 
the appearance of the appropriate record in the central database. 
Since the EOS system is designed for long-term storage, PyBARS first downloads raw data from 
the EOS storage to the local disk to speed up the processing. 
After the successful retrieval of the data to the local disk, the intermediate data files are removed for disk space economy, 
resulted data files are moved back to the central storage. 
To point back to the software version that produced the data, the latest git commit shortened hash 
is used in the directory hierarchy of resulted data. 
The results of each processing step are put into the central database as well as additional metadata such as run status, 
and are shown on the dashboard web service. 
DQM produces histograms categorised by different clusters, sections and channels archived in ROOT format. 
Web based data monitor is being developed to display plots produced by DQM.
Resulted data and produced alerts are obtained after about 2 hours from the moment when all 
raw files of typical run come to the central storage. 
For the active period with increased chemiluminescent lake background due to relatively soft trigger 
condition, this delay can increase by several times.
The physics production uses the validated calibration constants and the latest most precise  reconstruction algorithms. 
The physics production takes place on demand several times in a year according to the requirement of the physics analysis.
Each OM contains the controller and sensors that measure PMT pulse rate, high voltage, 
inside and outside temperatures, humidity, OM orientation, etc.
These data are also tranferred to the central storage and is put into central database in order to 
show time dependent plots of these characteristics on the dashboard with a latency of about 15 min.

\section*{Summary}

Baikal-GVD is gigaton-scale high-energy neutrino experiment under construction deep underwater in lake Baikal. 
The quasi online processing system has been developed to transform the raw data and make ready for physics analysis, 
as well as to produce alerts.
It also provides quick data transfer from the site and 
full data processing chain including data quality monitoring and event reconstructions.

\section*{Acknowledgements}

The work was partially supported by RFBR grants 20-02-00400. 
The CTU group acknowledges the support by European Regional
Development Fund-Project No. CZ.02.1.01/0.0/0.0/16 019/0000766.
We also acknowledge the technical support of JINR
staff for the computing facilities (JINR cloud).

\end{document}